\documentclass[aps,prl,twocolumn,showpacs,groupedaddress,longbibliography]{revtex4}

\usepackage[T1]{fontenc}
\usepackage[english]{babel}
\usepackage{epsfig}
\usepackage{graphicx}
\usepackage{color}
\usepackage{amsmath}
\usepackage{bbold}
\usepackage{float}
\usepackage{subfigure}
\usepackage{xifthen}
\usepackage{amsbsy}

\newcommand{\bs}[1] {\mathbf{#1}}

\newcommand{\up}{\uparrow}
\newcommand{\down}{\downarrow}

\newcommand{\chiRPA}[1]{\chi^{\mathrm{RPA}}_{#1}}
\newcommand{\chiSUSC}[2]{\chi^{ #2 }_{#1}}

\newcommand{\chiDMFT}[2] {
	\ifthenelse{\isempty{#2}}
	{ \chi_{ #1 } }
	{ \chi^{ #2 }_{ #1 } }
}
\newcommand{\chiBUBLOC}[2]{
	\ifthenelse{\isempty{#2}}
	{ \chi^{\mathrm{0,loc}}_{#1} }
	{ \chi^{\mathrm{0,loc}, #2 }_{#1} }
}
\newcommand{\chiBUB}[2]{
    \ifthenelse{\isempty{#2}}
    { \chi^{\mathrm{0} }_{#1}  }
    { \chi^{\mathrm{0}, #2 }_{#1} }
}
\newcommand{\chiBARE}[2]{
	\ifthenelse{\isempty{#2}}
	{ \chi^{0,\mathrm{bare} }_{#1} }
	{ \chi^{0,\mathrm{bare}, #2}_{#1} }
}

\begin{document}

\title{Dynamically enhanced magnetic incommensurability: \\ 
Effects of local dynamics on non-local spin-correlations in a strongly correlated metal}

\author {Demetrio~Vilardi}
\author{Ciro~Taranto}
\author{Walter~Metzner}
\affiliation{Max Planck Institute for Solid State Research, Heisenbergstrasse 1, D-70569 Stuttgart, Germany}

\date{\today}

%
%%%%%%%%%%%%%%%%%%%%%%%%%%%%%%%%%%%%%%%%%%%%%%%%%%%%%%%%%%%%%%%%%%%%%%%%%%%%%%%%%%%%%%%%%%%%%%
%
% ABSTRACT
%
%
%%%%%%%%%%%%%%%%%%%%%%%%%%%%%%%%%%%%%%%%%%%%%%%%%%%%%%%%%%%%%%%%%%%%%%%%%%%%%%%%%%%%%%%%%%%%%%
\begin{abstract}
We compute the spin susceptibility of the two-dimensional Hubbard model away from half-filling, and analyze the impact of frequency dependent vertex corrections as obtained from the dynamical mean field theory (DMFT). We find that the local dynamics captured by the DMFT vertex strongly affects non-local spin correlations, and thus the momentum dependence of the spin susceptibility. While the widely used random phase approximation yields commensurate N\'eel-type antiferromagnetism as the dominant instability over a wide doping range, the vertex corrections favor incommensurate ordering wave vectors away from $(\pi,\pi)$.
Our results indicate that the connection between the magnetic ordering wave vector and the Fermi surface geometry, familiar for weakly interacting systems, can hold in a strongly correlated metal, too.
\end{abstract}

%\pacs{}
\maketitle

%
%%%%%%%%%%%%%%%%%%%%%%%%%%%%%%%%%%%%%%%%%%%%%%%%%%%%%%%%%%%%%%%%%%%%%%%%%%%%%%%%%%%%%%%%%%%%%%
%
% INTRODUCTION
%
%
%%%%%%%%%%%%%%%%%%%%%%%%%%%%%%%%%%%%%%%%%%%%%%%%%%%%%%%%%%%%%%%%%%%%%%%%%%%%%%%%%%%%%%%%%%%%%%

{\it Introduction.---}
The dynamical mean field theory (DMFT) is one of the most successful tools to investigate strong correlations in interacting fermion systems, by means of a non-perturbative treatment of local dynamical correlations \cite{Metzner1989,Georges1991,Georges1996}.
The DMFT has been extended to include short-ranged non-local correlations via the dynamical cluster approximation \cite{Maier2005,Gull2013}, and long-ranged correlations by
diagrammatic approaches \cite{Rohringer2017,Toschi2007,Taranto2014}.
In combination with density functional theory, the DMFT provides an \emph{ab-initio} method for the calculation of electronic properties of real materials with strongly correlated electrons \cite{Anisimov1997,Kotliar2006}.

The DMFT self-consistency loop involves the calculation of one-particle quantities only. Instead, many physical observables such as magnetic properties and collective excitation spectra require the explicit calculation of two-particle quantities. The DMFT two-particle vertex \cite{Rohringer2012} is also a crucial ingredient for diagrammatic extensions of DMFT \cite{Rohringer2017}.

Since the calculation of the DMFT vertex is computationally demanding, susceptibilities are often computed by a random phase approximation (RPA) with DMFT propagators. This approach has frequently been applied to real materials, for example, to iron systems \cite{Igoshev2013}.
The importance of vertex corrections for the frequency dependence of the \emph{local} spin susceptibility has already been explicitly emphasized, 
e.g., in the context of realistic DMFT calculations for iron pnictides~\cite{Toschi2012,Liu2012}.
The effect of the vertex corrections on charge correlations has been studied with a more fundamental perspective by Hafermann \emph{et al.} \cite{Hafermann2012}, who focused on collective charge modes and gauge invariance.
In this work we reveal the effect of the local DMFT vertex on \emph{non-local} spin correlations, leading to qualitative changes of the momentum dependence of the spin susceptibility.

Using the two-dimensional Hubbard model as a test system, we show that the vertex corrections do not only affect the temperature scale at which strong magnetic fluctuations set in, but also the wave vector of the dominant magnetic instability.
Although local in space, the DMFT vertex strongly affects the non-local spin correlations.
Via its frequency dependence it drastically alters the momentum dependence of the susceptibility as compared to the momentum dependence of the RPA susceptibility, where the particle-hole bubble is dressed by the self-energy only.
In large parts of the phase diagram the RPA susceptibility is maximal at a wave vector $(\pi,\pi)$, pointing toward N\'eel-type commensurate antiferromagnetic order, while the susceptibility computed with vertex corrections exhibits pronounced maxima at incommensurate wave vectors on the Brillouin zone boundary away from $(\pi,\pi)$.

%
%%%%%%%%%%%%%%%%%%%%%%%%%%%%%%%%%%%%%%%%%%%%%%%%%%%%%%%%%%%%%%%%%%%%%%%%%%%%%%%%%%%%%%%%%%%%%%
%
% FORMALISM
%
%%%%%%%%%%%%%%%%%%%%%%%%%%%%%%%%%%%%%%%%%%%%%%%%%%%%%%%%%%%%%%%%%%%%%%%%%%%%%%%%%%%%%%%%%%%%%%
%

{\it Model.---}
The Hubbard model \cite{Montorsi1992} describes spin-$\frac{1}{2}$ lattice fermions with a purely local interaction. In standard second quantized notation, the Hamiltonian reads
\begin{equation}
 \mathcal{H} = \sum_{j,j',\sigma} t_{jj'} c^{\dagger}_{j,\sigma} c_{j',\sigma}
 + U \sum_{j} n_{j,\uparrow} n_{j,\downarrow} ,
\end{equation}
where $j$ and $j'$ are lattice indices, and $\sigma$ ($\uparrow$ or $\downarrow$) is the spin orientation. In applications to electrons in solids the interaction is repulsive, that is, $U>0$. We choose a two-dimensional square lattice and restrict the hopping amplitudes to
$t_{jj'} = -t$ for nearest neighbors and $t_{jj'} = -t'$ for next-to-nearest neighbors. 
Fourier transforming this hopping matrix yields the bare dispersion relation
\begin{equation}
 \varepsilon_{\mathbf{k}} =
 -2t \left( \cos{k_x} + \cos{k_y} \right) -4 t' \cos{k_x} \cos{k_y} .
\end{equation}
The Hubbard model on the square lattice has been proposed as a minimal model for the valence electrons in cuprate high-$T_c$ superconductors \cite{Anderson87}.

{\it Method.---}
To access the strongly interacting regime we use the DMFT, which captures non-perturbative effects such as the Mott metal-insulator transition \cite{Georges1996}. In DMFT the lattice model is mapped onto the Anderson impurity model (AIM), whose propagator is related to the lattice propagator by the self-consistency condition,
\begin{equation} 
% G^{\mathrm{loc}}_{\nu} = 
	\int \frac{d^2\bs{k}}{(2\pi)^2} 
	\frac{1}{i\nu +\mu-\epsilon_{\mathbf{k}}-\Sigma^{\mathrm{dmft}}_\nu } = 
	\frac{1}{\mathcal{G}^{-1}_\nu-\Sigma^{\mathrm{dmft}}_\nu} ,
\end{equation}
where $\nu$ is a fermionic Matsubara frequency, and $\mu$ is the chemical potential.
The lattice propagator on the left hand side is computed under the assumption that the self-energy is local and equals the impurity self-energy. $\mathcal{G}$ is the bare propagator of the AIM, and $\Sigma^{\mathrm{dmft}}$ is the self-energy obtained 
by solving the AIM associated with $\mathcal{G}$ and $U$. 

DMFT is exact in the limit of infinite lattice dimensions \cite{Metzner1989}. In finite dimensions it can be viewed as a local approximation for the self-energy. The momentum-dependent propagator is given by the Dyson equation,
\begin{equation}
\label{eq:Dyson}
G_{\nu,\bs{k}}=[i\nu-\epsilon_\mathbf{k}+\mu-\Sigma^{\mathrm{dmft}}_\nu]^{-1}.  
\end{equation}
A basic building block for the calculation of the spin susceptibility 
is the particle-hole pair propagator
\begin{equation} 
  \chiBUB{\omega,\bs{q}}{\nu,\bs{k}} = - G_{\nu,\bs{k}} G_{\nu+\omega,\bs{k}+\bs{q}} .
\end{equation}
From this quantity and the two-particle irreducible vertex in the magnetic channel,
$\Gamma^{\nu,\nu',\bs{k},\bs{k}'}_{\omega,\bs{q}}$, one can compute the spin susceptibility as $\chiSUSC{\omega,\bs{q}}{} = \int_{\nu,\bs{k}} \chiSUSC{\omega,\bs{q}}{\nu,\bs{k}}$, where
$\int_{\nu,\bs{k}} = T \sum_{\nu}\int \frac{d^2 \bs{k}}{(2\pi)^2}$, and
$\chiSUSC{\omega,\bs{q}}{\nu,\bs{k}}$ is determined by the linear integral equation
\begin{align}
  \chiSUSC{\omega,\bs{q}}{\nu,\bs{k}} &= \chiBUB{\omega,\bs{q}}{\nu,\bs{k}} - 
      \chiBUB{\omega,\bs{q}}{\nu,\bs{k}} \int_{\nu',\bs{k'}} 
          \Gamma^{\nu,\nu',\bs{k},\bs{k}'}_{\omega,\bs{q}}
       \chiSUSC{\omega,\bs{q}}{\nu',\bs{k}'} .
\label{eq:chiFULL} 
\end{align}
For the vertex we follow the notation of Rohringer \emph{et al.} \cite{Rohringer2012}, and $\Gamma$ corresponds to the antisymmetric spin combination
$\Gamma_{\up \up \up \up } - \Gamma_{\up \down \up \down}$.

Replacing the irreducible vertex with its lowest order in perturbation theory,
$\Gamma^{\nu,\nu',\bs{k},\bs{k}'}_{\omega,\bs{q}} = -U$, 
we obtain the RPA formula for the susceptibility,
\begin{equation}
\label{eq:RPAstandard}
 \chiRPA{\omega,\bs{q}} = \frac{\chiBUB{\omega,\bs{q}}{}}{1-U\chiBUB{\omega,\bs{q}}{}} ,
\end{equation}
where $\chiBUB{\omega,\bs{q}}{} = \int_{\nu,\bs{k}} \chiBUB{\omega,\bs{q}}{\nu,\bs{k}}$ is the polarization function, also known as particle-hole bubble.
The symmetric phase is stable only if the denominator in Eq.~(\ref{eq:RPAstandard}) is positive. A vanishing denominator for $\omega = 0$ at a certain wave vector $\bs{q}$ signals a magnetic instability. The structure of Eq.~(\ref{eq:RPAstandard}) implies that the maximum 
of the RPA-susceptibility $\chiRPA{}{}$ in momentum space coincides with the maximum of the particle-hole bubble $\chiBUB{}{}$.

A much better approximation for Eq.~(\ref{eq:chiFULL}) is achieved by substituting the two-particle irreducible vertex with the local counterpart $\Gamma^{\mathrm{dmft}}$ calculated 
for the effective AIM from the DMFT self-consistency loop, leading to
\begin{equation}
 \chiDMFT{\omega,\bs{q}}{\nu} = \chiBUB{\omega,\bs{q}}{\nu} - 
      \chiBUB{\omega,\bs{q}}{\nu} T \sum_{\nu'} 
          \Gamma^{\mathrm{dmft},\nu,\nu'}_{\omega}
       \chiDMFT{\omega,\bs{q}}{\nu'} ,
\label{eq:chiDMFTtmp}
\end{equation}
where $\chiBUB{\omega,\bs{q}}{\nu} =
 \int \frac{d^2\bs{k}}{(2\pi)^2} \chiBUB{\omega,\bs{q}}{\nu,\bs{k}}$.
This equation can be formally solved by a matrix inversion in Matsubara frequency space,
\begin{equation} 
  \chiDMFT{\omega,\bs{q}}{} = 
   T \sum_{\nu,\nu'} (D^{-1}_{\omega,\bs{q}})^{\nu,\nu'} \chiBUB{\omega,\bs{q}}{\nu'} ,
\label{eq:chiDMFT}
\end{equation}
where $D^{\nu,\nu'}_{\omega,\bs{q}} = \delta_{\nu,\nu'} 
 + T \chiBUB{\omega,\bs{q}}{\nu} \Gamma^{\mathrm{dmft},\nu,\nu'}_{\omega}$.

The susceptibility in Eq.~(\ref{eq:chiDMFT}) includes all the local correlations captured by the DMFT, while non-local correlations are treated by the ladder approximation. 
Note that $\chiBUB{}{}$ appearing in Eq.~(\ref{eq:chiDMFT}) already includes local correlations at the one-particle level due to the DMFT self-energy.
Since the vertex in Eq.~(\ref{eq:chiDMFT}) is local, the momentum-dependence of the susceptibility is generated by the particle-hole propagator. However, we will now see that, due to the convolution with the frequency dependent DMFT vertex, the momentum-dependence of the susceptibility does not simply trace the momentum-dependence of the bubble as in the RPA.

%
%%%%%%%%%%%%%%%%%%%%%%%%%%%%%%%%%%%%%%%%%%%%%%%%%%%%%%%%%%%%%%%%%%%%%%%%%%%%%%%%%%%%%%%%
%
% RESULTS
%
%%%%%%%%%%%%%%%%%%%%%%%%%%%%%%%%%%%%%%%%%%%%%%%%%%%%%%%%%%%%%%%%%%%%%%%%%%%%%%%%%%%%%%%%
%

%
\begin{figure}
\includegraphics[width=0.48\textwidth]{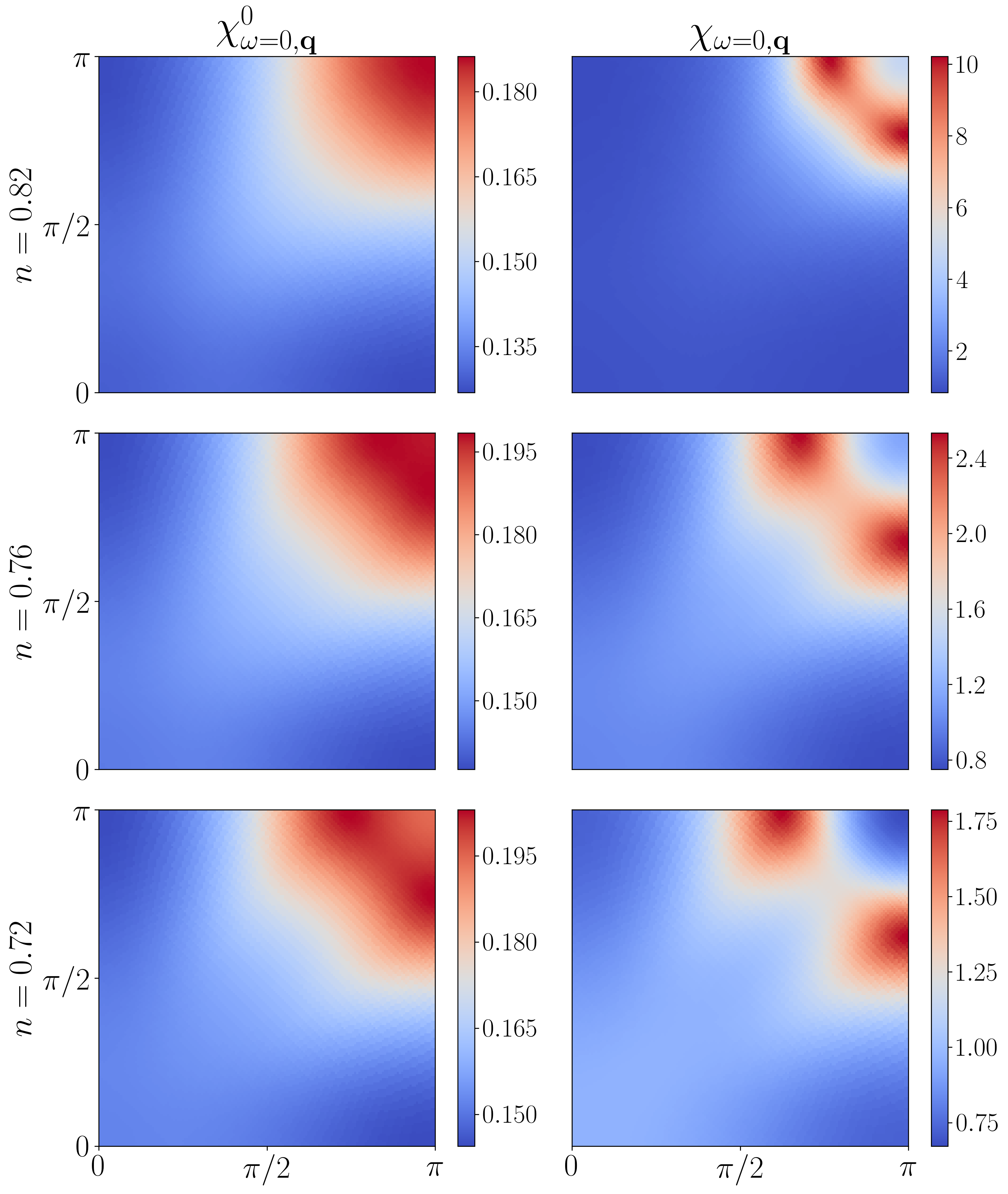}
\caption{Static particle-hole bubble (left) and static susceptibility (right) as a function of momentum in the first quadrant of the BZ. From top to bottom results for various densities are shown: $n=0.82$, $n=0.76$ and $n=0.72$. The other parameters are $U=8t$, $t'=-0.2t$, and $T=0.08t$.}
\label{fig:chiBZ}
\end{figure}
\begin{figure*}
\includegraphics[width=0.96 \textwidth]{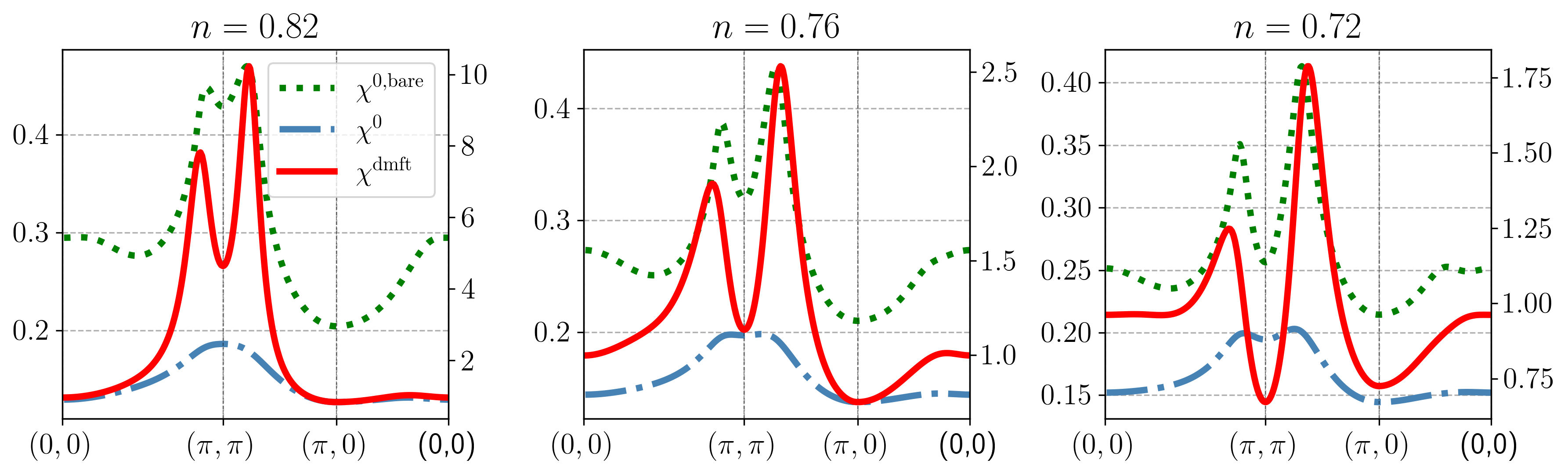}
\caption{Bare bubble (dotted line), DMFT bubble (dot-dashed line) and DMFT susceptibility (continuous line) plotted along a specific path in the BZ for $U=8t$, $t'=-0.2t$, and $T=0.08t$. From left to right: $n=0.82$, $n=0.76$ and $n=0.72$. The numbers on the left y-axes refer to both bare and DMFT bubble, while the numbers on the right y-axes refer to the susceptibility.}
\label{fig:pathbz005}
\end{figure*}
{\it Results.---}
In Fig.~\ref{fig:chiBZ} we show results for the static particle-hole bubble (left column) and the static DMFT susceptibility (right column) as a function of momentum in the first quadrant of the Brillouin zone (BZ) for various fermion densities $n < 1$. 
All quantities are computed for $t=1$. 
The interaction is rather strong ($U=8t$), and the temperature $T=0.08t$ has been chosen within the paramagnetic regime, that is, above the critical temperature \cite{footnote1} for a magnetic instability (within DMFT). 
One can clearly see that the positions of the maxima of the bubble and the susceptibility are generally distinct; in particular, for $n=0.82$, the maximum of the bubble is located at 
$\bs{q}=(\pi,\pi)$, while the DMFT susceptibility shows maxima for incommensurate vectors $\bs{q}=(\pi,\pi-2\pi \eta)$ and $\bs{q}=(\pi-2\pi\eta,\pi)$, with $\eta\approx0.12$.

Hence, for $n=0.82$, the widely used RPA formula~(\ref{eq:RPAstandard}) yields dominant commensurate 
antiferromagnetic correlations, since the momentum dependence of the RPA susceptibility is entirely determined by the particle-hole bubble. In this approximation the local correlations are taken into account only at the one-particle level, through the inclusion of the self-energy.
The behavior changes drastically when the local fluctuations are considered also at the two-particle level by including the DMFT vertex. The results for the susceptibility in the right panel of Fig.~\ref{fig:chiBZ} exhibit dominant incommensurate spin correlations for all shown densities. 
For $n=0.76$, both the particle-hole bubble and the susceptibility have incommensurate maxima, but at different positions. Reducing the filling further to $n=0.72$, the momentum $(\pi,\pi)$ becomes a marked local minimum for the bubble and even a global minimum for the DMFT susceptibility. 

The peak structure of the bubble and the susceptibility can be seen more clearly in a plot along the $\mathrm{\Gamma}$-M-X-$\mathrm{\Gamma}$ path in the BZ, as shown in Fig.~\ref{fig:pathbz005}. Here, it is evident that $(\pi,\pi)$ becomes a global minimum for the susceptibility at filling $n=0.72$. In this plot we also show the \emph{bare} bubble $\chiBARE{}{}$, which is computed without self-energy feedback and with the bare chemical potential. 

Despite the fact that the bare bubble does not enter in Eq.~(\ref{eq:chiFULL}), since $\chi^0$ is evaluated with self-energy feedback, the momentum dependence of the DMFT susceptibility resembles much more the one of the bare bubble rather than the bubble with dressed propagator. This is remarkable since the self-energy strongly affects the particle-hole bubble: first, as expected, the self-energy globally suppresses the bubble; second, and more importantly, it smears the peak in momentum space and thus reduces or even eliminates the shift $\eta$.
By contrast, the two-particle vertex has the opposite effect: it sharpens the peak and increases $\eta$.

\begin{figure} 
\includegraphics[width=0.45 \textwidth]{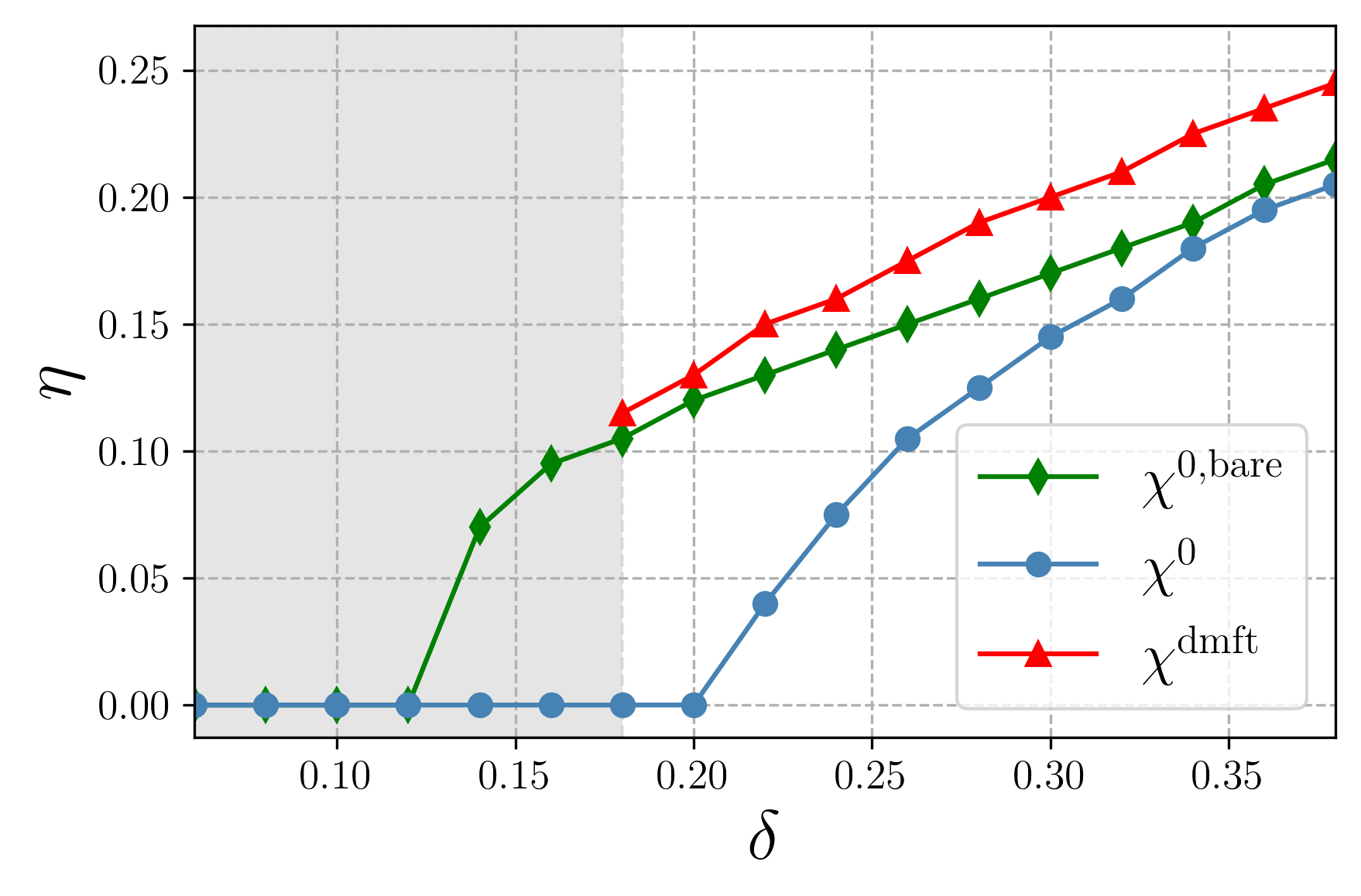}
\caption{Incommensurability $\eta$ as a function of the doping $\delta=1-n$, for $U=8t$, $t'=-0.2t$, and $T=0.08t$. The different curves refer to the DMFT susceptibility, the DMFT bubble with self-energy, and the bare bubble, respectively. The grey area indicates the doping values where Eq.~(\ref{eq:chiDMFTtmp}) has no solution, due to a magnetic instability in that regime.}
\label{fig:eta}
\end{figure}
\begin{figure*} 
 \includegraphics[width=0.9 \textwidth]{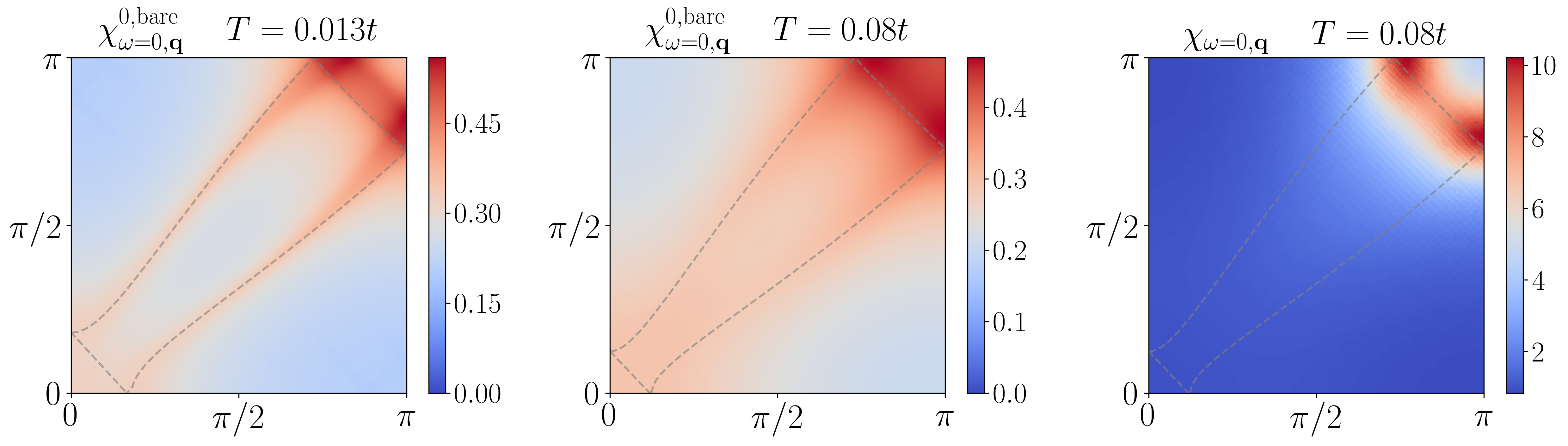}
 \includegraphics[width=0.9 \textwidth]{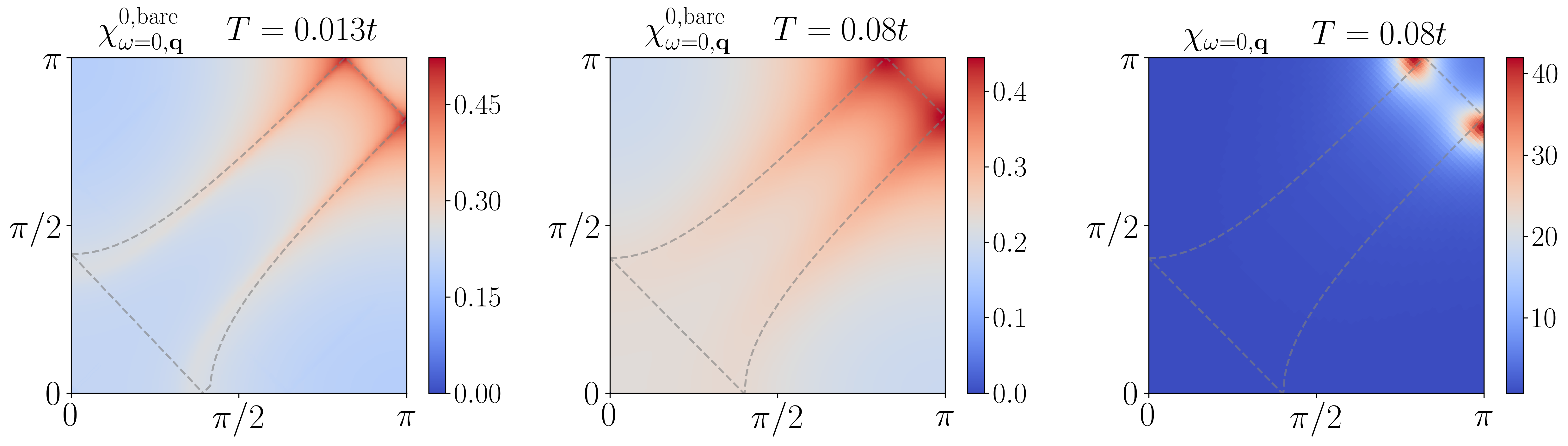}
 \caption{From left to right, bare bubble at $T=0.013t$, bare bubble at $T=0.08t$, and susceptibility at $T=0.08t$ as a function of momentum for $U=8t$ and $n=0.82$. Top: $t'=-0.2t$, bottom: $t'=-0.08t$. The dashed lines represent the nesting vectors of the Fermi surface as explained in Ref.~\onlinecite{Holder2012}.}
 \label{fig:susc_lowT_comparison}
\end{figure*}
To study further the relation between the particle-hole bubble and the DMFT susceptibility, in Fig.~\ref{fig:eta} we show the corresponding incommensurabilities $\eta$ as function of the doping $\delta=1-n$.
The maximum of the dressed particle-hole bubble stays at $(\pi,\pi)$ for doping smaller than $0.2$, and moves away from $(\pi,\pi)$ only for $\delta > 0.2$.
On the other hand, the incommensurability vector of the DMFT susceptibility is finite 
already for doping $x=0.14$, and is always larger than the one of the particle-hole 
bubble. By contrast, in the paramagnetic regime where the DMFT susceptibility is well defined, the incommensurability vector of $\chiDMFT{}{}$ is quite close to the one obtained from the \emph{bare} bubble.
When the doping is increased, the difference between the incommensurability of the dressed bubble and of the susceptibility is gradually reduced.
Similar results are obtained for $t'=-0.08t$.

The position of the peaks of the bare particle-hole bubble at low temperatures is determined by crossing points of \emph{nesting-lines} (or ''$2k_F$-lines'') in the BZ (see, for example, Ref.~\cite{Holder2012}). The latter are lines formed by the set of all nesting vectors of the Fermi surface, which connect Fermi momenta with collinear Fermi velocities. The smearing effect of the self-energy on the bubble spoils the connection with the Fermi surface geometry.
Since the vertex correction strongly affects the momentum dependence of the susceptibility, the question arises whether the vertex restores a connection between the susceptibility and the Fermi surface.

To further investigate this point, in Fig.~\ref{fig:susc_lowT_comparison} we plot the susceptibility at $T=0.08t$, already shown in previous plots, together with the bare bubble at the same temperature, and also at a lower temperature $T=0.013t$, where the signature of the Fermi surface is more pronounced.
Structures along the nesting-lines parallel to the BZ diagonals are visible only in the bubble, not in the susceptibility. However, the positions of the incommensurate peaks near the crossing points of nesting lines on the BZ boundary are quite similar in both quantities. 
This similarity suggests a connection between the peaks in the DMFT susceptibility and the Fermi surface geometry. 

%
%%%%%%%%%%%%%%%%%%%%%%%%%%%%%%%%%%%%%%%%%%%%%%%%%%%%%%%%%%%%%%%%%%%%%%%%%%%%%%%%%%%%%%%%
%
% CONCLUSIONS
%
%%%%%%%%%%%%%%%%%%%%%%%%%%%%%%%%%%%%%%%%%%%%%%%%%%%%%%%%%%%%%%%%%%%%%%%%%%%%%%%%%%%%%%%%
%

{\it Conclusion.---}
In summary, we have shown that the momentum dependence of the spin susceptibility in a strongly interacting Fermi system is drastically affected by the local two-particle dynamics. While the susceptibility computed within the DMFT without vertex correction yields commensurate N\'eel order as the dominant magnetic instability over a wide density range below half-filling, the local but strongly frequency dependent DMFT vertex shifts the ordering wave vector away from $(\pi,\pi)$ toward generally incommensurate wave vectors of the form $(\pi,\pi - 2\pi\eta)$. Similar conclusions apply also for the three-dimensional Hubbard model~\cite{Schaefer2018}.
The position of the peaks of the DMFT susceptibility with vertex correction in the BZ is strikingly close to the peaks obtained from the \emph{bare} particle-hole bubble, which is entirely determined by the Fermi surface geometry. This is remarkable since the self-energy in a strongly interacting system leads to a strong life-time broadening, which blurres the Fermi surface structures. Obviously, the vertex correction partially cancels this self-energy effect and restores the connection to the Fermi surface geometry. 

Strong incommensurate magnetic correlations are observed in cuprate superconductors and other layered transition metal compounds.
They also emerged in theoretical studies of the two-dimensional $tJ$-model, that is, the strong coupling limit of the Hubbard model \cite{Shraiman1989,Kotov2004}.
Clearly, the DMFT fails to capture many important aspects of such low-dimensional systems. However, the DMFT results show that the local two-particle dynamics is of crucial importance for the non-local spin correlations in a strongly interacting Fermi system.

We are grateful to S.~Andergassen, P.~Hansmann, G.~Rohringer, T.~Sch\"afer, and A.~Toschi for valuable discussions.

%\vspace*{5mm}


\begin{thebibliography}{2}
 \bibitem{Metzner1989} W.~Metzner and D.~Vollhardt,
 \emph{Correlated Lattice Fermions in $d=\infty$ Dimensions}, Phys. Rev. Lett. \textbf{62}, 1066 (1989).
 
 \bibitem{Georges1991} A.~Georges and G.~Kotliar,
 \emph{Hubbard Model in Infinite Dimensions}, Phys. Rev. B \textbf{45}, 6479 (1992).
 
 \bibitem{Georges1996} A.~Georges, G.~Kotliar, W.~Krauth, and M.~J.~Rozenberg, 
 \emph{Dynamical mean-field theory of strongly correlated fermion systems and the limit of infinite dimensions}, Rev. Mod. Phys. \textbf{68}, 13-125 (1996).
 
 \bibitem{Maier2005} T.~Maier, M.~Jarrell, T.~Pruschke, and M.H.~Hettler, 
 \emph{Quantum cluster theories}, Rev. Mod. Phys. \textbf{77}, 1027 (2005).
		
 \bibitem{Gull2013} E.~Gull, O.~Parcollet, and A.J.~Millis,
 \emph{Superconductivity and the Pseudogap in the Two-Dimensional Hubbard Model}, Phys. Rev. Lett. \textbf{110}, 216405 (2013).
 
 \bibitem{Rohringer2017} G.~Rohringer, H.~Hafermann, A.~Toschi, A.A.~Katanin, A.E.~Antipov, M.I.~Katsnelson, A.I.~Lichtenstein, A.N.~Rubtsov, and K.~Held, 
 \emph{Diagrammatic routes to non-local correlations beyond dynamical mean field theory}, arXiv:1705.00024.
 
 \bibitem{Toschi2007} A.~Toschi, A.A.~Katanin, and K.~Held,
 \emph{Dynamical vertex approximation: A step beyond dynamical mean-field theory}, Phys. Rev. B \textbf{75}, 045118 (2007).
 
 \bibitem{Taranto2014} C.~Taranto, S.~Andergassen, J.~Bauer, K.~Held, A.~Katanin, W.~Metzner, G.~Rohringer, and A.~Toschi, 
 \emph{From Infinite to Two Dimensions through the Functional Renormalization Group}, Phys. Rev. Lett. \textbf{112}, 196402 (2014).
 
 \bibitem{Anisimov1997} V.I.~Anisimov, A.I.~Poteryaev, M.A.~Korotin, A.O.~Anokhin, and G.~Kotliar, 
 \emph{First-principles calculations of the electronic structure and spectra of strongly correlated systems: dynamical mean-field theory}, Journal of Physics: Condensed Matter \textbf{9}, 7359 (1997).
   
 \bibitem{Kotliar2006}  G.~Kotliar, S.Y.~Savrasov, K.~Haule, V.S.~Oudovenko, O.~Parcollet, and C.A.~Marianetti, 
 \emph{Electronic structure calculations with dynamical mean-field theory}, Rev. Mod. Phys. \textbf{78}, 865 (2006).

 \bibitem{Rohringer2012} G.~Rohringer, A.~Valli, and A.~Toschi, 
 \emph{Local electronic correlation at the two-particle level}, Phys. Rev. B \textbf{86}, 125114 (2012).
 
 \bibitem{Igoshev2013} P.A.~Igoshev, A.V.~Efremov, A.I.~Poteryaev, A.A.~Katanin, and V.I.~Anisimov, 
 \emph{Magnetic fluctuations and effective magnetic moments in $\gamma$-iron due to electronic structure peculiarities}, Phys. Rev. B \textbf{88}, 155120 (2013).

 \bibitem{Toschi2012} A.~Toschi, R.~Arita, P.~Hansmann, G.~Sangiovanni, and K.~Held,
 \emph{Quantum dynamical screening of the local magnetic moment in Fe-based superconductors}, Phys. Rev. B \textbf{86}, 064411 (2012).

 \bibitem{Liu2012} M.~Liu, L.W.~Harriger, H.~Luo, M.~Wang, R.A.~Ewings, T.~Guidi, H.~Park, K.~Haule, G.~Kotliar, S.M.~Hayden, and P.~Dai, 
 \emph{Nature of magnetic excitations in superconducting BaFe$_{1.9}$Ni$_{0.1}$As$_{2}$}, Nature Physics {\bf 8}, 376-381 (2012)
 
 \bibitem{Hafermann2012} H.~Hafermann, E.G.C.P.~van Loon, M.I.~Katsnelson, A.I.~Lichtenstein, and O.~Parcollet, 
 \emph{Collective charge excitations of strongly correlated electrons, vertex corrections, and gauge invariance}, Phys. Rev. B \textbf{90}, 235105 (2014).
 
 \bibitem{Montorsi1992} A.~Montorsi,
 \emph{The Hubbard Model: A Reprint Volume} (World Scientific, 1992). 
 
 \bibitem{Anderson87} P.W.~Anderson,
 \emph{The Resonating Valence Bond State in La2CuO4 and Superconductivity}, Science {\bf 235}, 1196 (1987).
 
 \bibitem{footnote1} Note that the DMFT yields a finite magnetic transition temperature in any dimension. In applications to two-dimensional systems the DMFT thus violates the Mermin-Wagner theorem due to the lack of non-local order parameter fluctuations.
 
 \bibitem{Holder2012} T.~Holder and W.~Metzner, 
 \emph{Incommensurate nematic fluctuations in two-dimensional metals},
 Phys. Rev. B \textbf{85}, 165130 (2012). 
 
 \bibitem{Schaefer2018} We thank Thomas Sch\"afer for confirming this by applying the
 ladder version of the dynamical vertex approximation as used for the calculation of
 spin correlations in T.~Sch\"afer, A.A.~Katanin, K.~Held, and A.~Toschi,
 \emph{Interplay of Correlations and Kohn Anomalies in Three Dimensions: Quantum
 Criticality with a Twist}, Phys. Rev. Lett. \textbf{119}, 046402 (2017).
 
 \bibitem{Shraiman1989} B.I.~Shraiman and E.D.~Siggia,
 \emph{Spiral phase of a doped quantum antiferromagnet},
 Phys. Rev. Lett. \textbf{62}, 1564 (1989).
  
 \bibitem{Kotov2004} V.N.~Kotov and O.P.~Sushkov,
 \emph{Stability of the spiral phase in the two-dimensional extended t-J model},
 Phys. Rev. B \textbf{70}, 195105 (2004).
 
\end{thebibliography}
\end{document}